  \documentclass[preprint2]{aastex}

\shortauthors{New \& Shapiro}
\shorttitle{Evolution of Differentially Rotating Supermassive Stars}

\begin{document}


\title{Evolution of Differentially Rotating Supermassive Stars \\
to the Onset of Bar Instability}


\author{Kimberly C. B. New}
\affil{Los Alamos National Laboratory, Los Alamos, NM 87545}
\email{knew@lanl.gov}

\author{Stuart L. Shapiro\altaffilmark{1}}
\affil{Department of Physics, University of Illinois at Urbana-Champaign,
Urbana, IL 61801}
\email{shapiro@astro.physics.uiuc.edu}


\altaffiltext{1}{Department of Astronomy and National Center for
Supercomputing Applications, University of Illinois at Urbana-Champaign,
Urbana, IL 61801}


\begin{abstract}

Thermal emission from a rotating, supermassive star will cause 
the configuration to contract slowly and spin up.
If internal viscosity and magnetic fields are sufficiently weak, 
the contracting star will rotate differentially.  
For each of the six initial angular momentum distributions considered,
a cooling supermassive star will likely encounter the dynamical bar mode
instability, which may trigger the growth of nonaxisymmetric bars.
This scenario will lead to the generation of long wavelength
gravitational waves, which could be detectable by
future space-based laser interferometers like LISA. 
\bigskip

\end{abstract}




\bigskip

\section{Introduction}

The existence of supermassive black holes (SMBHs) and
the theory of their role as fuel sources for active galactic nuclei
and quasars are supported by a growing body of solid
observational evidence (see, e.g., the reviews of Rees 1998
and Macchetto 1999,  and references therein).  For example, VLBI
observations of the Keplerian motion of masing knots orbiting
the nucleus of NGC 4258 indicate the presence of a central object
of mass $\sim 3.6 \times 10^7 M_{\sun}$ and radius $< 13$ pc
\citep{miyo95}.  The most reasonable explanation is that this 
extremely compact object is a black hole.  High resolution HST
observations of matter orbiting the nuclei of M87 and NGC 4151
also show nearly Keplerian motion around SMBHs \citep{macc97,
wing99}.  In addition, reverberation mapping and photoionization
methods indicate the presence of SMBHs in the centers of 17 Seyfert
quasars \citep{wand99}.

However, the nature of the progenitors of SMBHs is rather uncertain
(see Rees 1984, 1999 for overviews).  SMBH formation scenarios
that involve the stellar dynamics of dense clusters and the hydrodynamics
of supermassive stars have both been proposed.  The suggested
stellar dynamical routes focus on the evolution of dense star
clusters.  These clusters are formed via a conglomeration of stars,
produced by fragmentation of the primordial gas.  
In one such cluster
scenario, massive stars form via stellar collisions and mergers 
in the cluster
and then evolve into stellar-mass black holes.  The merger of these
holes, as they grow and settle to the cluster center, leads to the
build up of one or more SMBHs (see, e.g., Quinlan and Shapiro 1990, 
Lee 1995, and
references therein).  Alternatively, dense clusters composed of compact
stars may become dynamically unstable to relativistic core collapse
and thus may give birth to SMBHs \citep{zepo65,shte85a,shte85b,qush87}.

The proposed hydrodynamical routes to SMBH formation center on the
evolution of supermassive stars (SMSs).  SMSs may contract directly
out of the primordial gas or they may develop from a supermassive
gas cloud that has been built up from the fragments of stellar
collisions in clusters (Sanders 1970; Begelman and Rees 1978;
Haehnelt, Natarajan, and Rees 1998; however, see also Loeb and Rasio 1994;
Abel, Bryan, and Norman 2000; Bromm, Coppi, and Larson 1999).
The evolution of SMSs will ultimately lead to relativistic dynamical collapse 
\citep{iben63,chan64a,chan64b} and thus possibly to the formation of SMBHs.

In addition to their relevance to AGNs and quasars, SMBHs and
their formation events are likely candidates
for detection by proposed space-based gravitational wave detectors,
like the Laser Interferometer Space Antenna (LISA) \citep{folk98}.  LISA would
be very sensitive to the long wavelength, low frequency gravitational
radiation that supermassive objects are expected to emit
\citep{thbr76,thor95}.  For example, LISA may be able to detect
the collapse of a SMS to a SMBH or the coalescence of two SMBHs
\citep{lisa95}.  The rates of such events are largely unknown due to 
the uncertainty in the formation mechanism of SMBHs.

Identifying the scenario by which SMBHs form is hence
of fundamental importance to a number of areas of astrophysics.
This work examines one of the
possible progenitors of SMBHs discussed above, SMSs. 
The structure and evolution of SMSs have been explored by many
authors (see, e.g., Zel'dovich and Novikov 1971 and
Shapiro and Teukolsky 1983, and references therein).
In this paper, we focus specifically on the quasistatic evolution of
SMSs as they contract and cool due to thermal emission. 

Two related factors that affect the cooling phase of a SMS
are the strengths of its viscosity and magnetic fields and
the nature of its angular
momentum distribution.  If strong enough, viscosity will enforce
uniform rotation throughout the star as it contracts.
However, the strength of
molecular viscosity is unlikely to be large enough
to accomplish this, so turbulent viscosity or magnetic fields must
provide the necessary dissipation \citep{zeno71, spru99a, spru99b, bash99}.
However, it is not known whether these agents are sufficiently strong
to maintain uniform rotation in SMSs as they evolve.

If a SMS is rotating prior to its contraction phase, conservation
of angular momentum requires the star to spin up and therefore
increase its ratio of kinetic energy $T$ to gravitational potential
energy $W$, $\beta=T/|W|$.  If viscosity or magnetic fields are
able to maintain
uniform rotation, the star will be spun up to its mass--shedding
limit $\beta_{shed} = 8.99 \times 10^{-3}$. The value is small because a SMS
is dominated by radiation pressure and thus is well represented by
an $n$=3 polytropic equation of state (see \S 2.1.1).  As a result,
it is quite centrally and cannot
support much rotation when spinning uniformly before mass at the equator
is driven into Keplerian orbit. Thereafter, it will evolve along a
mass--shedding sequence, losing both mass and angular momentum, and
will contract to the point of relativistic instability and collapse.
\citet{bash99} have recently investigated this evolutionary
scenario in detail and have described how it may lead to SMBH formation.
Specifically, they identified the critical configuration at the onset
of the relativistic radial instability.  They also demonstrated
that the nondimensional (in units where $G$=$c$=1) ratios
$T/|W|$; $R/M$, where $R$ is the radius and $M$ is the mass; and $J/M^2$,
where $J$ is the angular momentum, are universal numbers (i.e., they do
not depend on $M$, $R$, or $J$ for the crititcal configuration or on
the history of the star).

This work investigates the extreme opposite regime of SMS cooling evolution,
the low-viscosity  and low-magnetic field limits.  
In this case, as the star contracts and spins up, neither
viscosity nor magnetic fields will be able to enforce uniform rotation (even if
the star initially rotates uniformly).  The outcome
of the evolution will then depend on the star's initial angular
momentum distribution (or rotation law), as the star will conserve
its angular momentum on cylindrical shells as it contracts.
One possible outcome
is that the star will again be spun-up to the mass-shedding limit
(at a value of $\beta_{shed}$ higher than that for uniform rotation).
The subsequent evolution is not clear. It is likely that mass and
angular momentum will be ejected from the equator
as the star undergoes further
cooling. It is possible that it will continue to evolve in
quasistationary equilibrium 
until reaching the onset of dynamical collapse.
If it implodes to a SMBH, the matter is likely to emit an appreciable
burst of gravitational radiation just prior to entering the
event horizon.

An alternative outcome of low-viscosity, low-magnetic field
contraction of a differentially rotating SMS is that it will
encounter the {\it dynamical} ``bar mode'' instability
at a critical value $\beta_{bar}$ prior to reaching $\beta_{shed}$.
Such global rotational instabilities in fluids arise
from nonaxisymmetric modes, which vary like $e^{\pm im\phi}$, where $m$=2  is
the bar mode.  As the bar mode develops, angular momentum
and mass will be transported outwards \citep{tdm85,pick96,idp00,nct00,brow00}. 

If enough angular momentum
is removed, collapse to a SMBH could ensue.  
A SMS undergoing a bar mode distortion would be
a strong source of quasi-periodic gravitational radiation because
of its highly nonaxisymmetric structure. 
We note that a bar might form during catastrophic
collapse in any scenario, if $\beta (\propto 1/R)$
grows above the critical value for the dynamical bar mode
instability.

The outcome of the cooling and the resultant spin-up of
a low-viscosity SMS depends critically on its initial angular
momentum distribution.  However, the nature of the angular
momentum distribution of SMSs is largely unknown.
The purpose of this paper is to examine the evolution of SMS models
by assuming several different initial rotation laws to
determine, for example, if any of them might cause 
the star to encounter the dynamical bar mode,
prior to mass-shedding
and collapse
(i.e., to determine if $\beta_{bar}<\beta_{shed}$).

To make this determination, we analyze several 
Newtonian equilibrium sequences of SMS models.  The
models along each sequence are constrained to 
have the identical mass $M$,
angular momentum $J$, and angular momentum distribution  
but decreasing entropy, due to cooling. 
Such a sequence mimicks
the quasistatic evolutionary track of a SMS as it slowly cools and contracts.  
We find that along the sequence, the stars flatten as measured by the
ratio of polar radius
$R_p$ to equatorial radius $R_e$.
We examine each sequence to determine whether the limits  $\beta_{shed}$ and/or
$\beta_{bar}$ are reached.  Note that 3D hydrodynamical simulations are
necessary to determine precisely $\beta_{bar}$ for each sequence.
However, previous linear and nonlinear stability analyses
indicate that $0.14\lesssim\beta_{bar}
\lesssim 0.27
$ \citep{chan69,shte83,duto85,imam85,mana85,
hach88,toha90,pick96,toma98,idp00}.
See \S 4 for further discussion of these stability limits. 

We note that {\it secular} version of the bar instability, which sets
in at a lower value of $\beta$, is not relevant for the scenario
described here.  The reason for this is that the viscosity that could
drive the mode is assumed to be vanishingly small and, hence, the secular
timescale is too long \citep{bash99}.
The timescale for gravitational radiation to drive
the secular bar mode is also much longer than the evolution (cooling) timescale
\citep{chan70}.

In the low-viscosity and low-magnetic field limits, the angular momentum of mass
shells on concentric cylinders in a contracting star will be preserved
\citep{boos73,tass78}.
Thus we confine our discussion to
equilibrium sequences constructed with rotation laws that conserve
the angular momentum distribution in this manner.
The sequences examined here include one we construct below
with the $n'$=3 rotation law (see \S 2.2). This sequence 
represents the cooling evolution
of a SMS that is nearly spherical and rotating uniformly 
at formation prior to its contraction phase. It is the 
low-viscosity, low-field  analog of the uniformly rotating
sequence considered in 
Baumgarte \& Shapiro (1999); in the present case,
differential rotation will ensue once cooling gets underway.
Because the initial rotation profiles of SMSs are unknown, we
also examine alternative sequences constructed by Hachisu, Tohline, and
Eriguchi (1988; hereafter HTE)
that describe the evolution of SMSs with nonuniform
initial rotation laws.

The remainder of this paper is organized as follows.  In \S 2, we
describe the numerical methods used to construct our $n'$=3 equilibrium
sequence and the sequences constructed by HTE.
We present our $n'$=3 sequence and summarize the sequences
constructed by HTE in \S 3.  In \S 4, we outline possible
evolutionary scenarios for SMSs, based on the sequences described
in \S 3.  We discuss these results in \S 5.

\section{Numerical Methods for SMS Model Construction}

We have constructed an equilibrium sequence of SMS models,
with a rotation law appropriate for SMSs that rotate
uniformly prior to contraction.
The individual models along the sequence have the same
rotation law, $M$, and $J$, but decreasing entropy and axis ratios $R_p/R_e$.
Each sequence is a quasistatic approximation
to the evolution of a SMS that is contracting due to thermal
emission.  Modeling this phase of the evolution of SMSs
with such equilibrium sequences is appropriate as the
cooling timescale is longer than the hydrodynamic timescale
(for $M\lesssim10^{13}M_\sun$, Baumgarte and Shapiro 1999, and
references therein)

\subsection{The Self-Consistent Field Method}

The structure of a fluid rotating about the $z$ axis with constant angular
velocity $\Omega=\Omega(\varpi)$, where $\varpi$ is the distance
from the rotation axis, is described
by the following expression:
\begin{equation}
\frac{1}{\rho} \nabla P + \nabla \Phi + h_0^2 \nabla \Psi(\varpi)=0,
\end{equation}
where $P$ is the pressure, $\Phi$ is the gravitational potential,
$\Psi(\varpi)=-(1/h_0^2)\int \Omega^2(\varpi)\varpi\, d\varpi$
is the centrifugal potential, and $h_0$ is a constant.  Such a fluid
is said to be in hydrostatic equilibrium because the forces due to
its pressure and to its gravitational and centrifugal potentials
are in balance.

The method we have used to construct SMSs in hydrostatic equilibrium
is Hachisu's grid based, iterative, axisymmetric Self-Consistent Field
(HSCF) technique (Hachisu 1986, see also New 1996).

\subsubsection{Equation of state}
The HSCF method requires the fluid to have a barotropic equation of
state (EOS) $P=P(\rho)$.  All of the SMS models we have constructed
have a polytropic EOS for which
\begin{equation}
P=K\rho^{1+1/n},
\end{equation}
where $n$ is the polytropic index and $K$ is the entropy constant.
The enthalpy of a polytrope is
\begin{equation}
H=(n+1)K\rho^{1/n}.
\end{equation}

The structure of a SMS is well represented by an $n$=3 polytrope
(see \S 17.2 of Shapiro and Teukolsky 1983).
This is because SMSs are dominated by thermal
radiation pressure (for $M\gtrsim 10^6 M_\sun$, Zel'dovich and
Novikov 1971; Fuller, Woosley, and Weaver 1986)
and are convective with constant entropy per baryon 
throughout (see Loeb and Rasio 1994 for a simple proof of this property).
The value of $K$ is determined by the
entropy per baryon $s$ in the star and is given by
\begin{equation}
K=\frac{a}{3}\biggl(\frac{3s}{4m_Ha}\biggr)^{4/3},
\end{equation}
where $a$ is the radiation density constant and $m_H$ is the mass
of a hydrogen atom (see eq. 17.2.6 in Shapiro and Teukolsky 1983).
The constant $s$ can be expressed in terms
of the matter temperature $T_m$, baryon denisty $n_b$, and $a$:
\begin{equation}
s=\frac{4}{3} \frac{a T_{m}^3}{n_b}.
\end{equation}

For polytropic EOSs, the sequences constructed with
Hachisu's SCF method 
are parameterized solely by the axis ratio $R_p/R_e$.
The method requires that the grid cell locations of the two 
surface points, $A=R_e$ and $B=R_p$, be specified.

\subsubsection{Rotation Laws}

In the low-viscosity limit, the angular momentum of
a contracting star will be preserved on cylindrical mass shells
\citep{boos73,tass78}.  Thus the rotation laws used in sequences
representative of such evolution must enforce
this conservation.  The $n'$=3 law and each of the laws employed in the
sequences constructed by HTE are defined such that the specific angular
momentum profile $j(m)$ is identical for each model constructed.
Here $m$ is the mass interior to cylindrical radius $\varpi$.
Because $j$ is the angular momentum per unit mass and is constant
on cylindrical surfaces, a $j(m)$ law does ensure that the
angular momentum distribution on cylindrical mass shells will be 
preserved from model to model along the sequence.

\subsubsection{Converged Configurations}

The quality of the converged configurations can be estimated
in terms of the virial error $VE$.  The $VE$ provides a measure
of how well the energy is balanced and is defined as
\begin{equation}
VE\equiv 2T+W+3\int P \, dV,
\end{equation}
where V is the volume of the model. The quantity $VE$ is zero in
strict equilibrium.
The $VE$s for models on this sequence range from $\sim 10^{-2}$ (for
the very smallest axis ratios) to $\sim 10^{-6}$, where the individual terms
in equation (9) are of order unity (see Table 1).

Our axisymmetric equilibrium configurations were constructed on uniform,
$1024\times 1024$ $(\varpi,z)$ grids.  Equatorial symmetry through
the $z$=0 plane was assumed; this symmetry condition is valid
for systems with $\Omega(\varpi)$ and barotropic EOSs \citep{tass78}.

Note that initially we had difficulty obtaining converged models
on the $n$=$n'$=3
sequence (see \S 2.2, 3.1) with axis ratios $<0.400$.  For these models, the
HSCF iterations oscillated between two or more states without
converging.  In order to obtain converged models with these
axis ratios, we implemented a modification to the HSCF algorithm
suggested in \citet{pick96}.
Instead of using $\rho_i$ from iteration $i$ as input for the next iteration
step $i+1$, a combination of $\rho_i$ and the old density $\rho_{i-1}$ is used
(``under-relaxation''):
\begin{equation}
\rho^{'}_{i} = (1-\delta)\rho_i + \delta \rho_{i-1},
\end{equation}
where $\delta$ is a constant (in this work, $0.75\leq\delta\leq 0.99$).

\subsection{Scale of Models}

As mentioned above,
polytropic sequences constructed with the HSCF code are
parameterized soley by the axis ratio $R_p/R_e$.
That is, for a given EOS and rotation law, the choice of
$R_p/R_e$ produces a unique model.
A dimensional scale (i.e., mass, radius, etc.) 
for this model may be chosen after its construction.
The individual
models along a sequence representative of the cooling
evolution of a SMS have the same $M$ and $J$.  As will be demonstrated
below, the choice of $M$ and $J$ for a sequence sets the dimensional
scale for its individual models.

The HSCF computations are carried
out in dimensionless form, with all quantities normalized
in terms of $G$, $R_e$, and $\rho_{max}$.  Here
$\rho_{max}$ is the maximum density of the model.
In HSCF units, $M$ and $J$ are normalized according to:
\begin{mathletters}
\begin{eqnarray}
\hat{M}&=&\frac{M}{R_e^3\rho_{max}}, \\
\hat{J}&=&\frac{J}{G^{\frac{1}{2}}R_e^5\rho_{max}^{\frac{3}{2}}}.
\end{eqnarray}
\end{mathletters}
$\!\!$Here, and in what follows, carets denote normalized quantities.

The HSCF method produces unique values
of the normalized quantites (e.g., $\hat{M}$, $\hat{J}$, etc.) for
each input axis ratio $R_p/R_e$.
After a converged model has been constructed, the normalized quantities
are known.
If physical values of $M$ and $J$ are then chosen to specify the sequence, 
equations (8) can be used to compute $R_e$ and $\rho_{max}$ for each
model along the sequence:
\begin{equation}
R_e=G^{-1}\left(\frac{J}{\hat{J}}\right)^2
\left({\frac{\hat{M}}{M}}\right)^3,
\end{equation}
\begin{equation}
\rho_{max}=\frac{M}{\hat{M}R_e^3}.
\end{equation} 

Once $R_e$ and $\rho_{max}$ are computed, they can be used
to convert normalized quantities to physical units.  For
example,
the definition of $H$ from equation (3) and its normalization,
$\hat{H}=H/(G R_e^2 \rho_{max})$, can be used to determine a physical
value for $K$:
\begin{equation}
K=\frac{1}{1+n} G \hat{H}_{max} R_e^2 \rho_{max}^{1-1/n} .
\end{equation}
Note that when $n$=3, equation (10) leads to the following
simplification of equation (11):
\begin{equation}
K=\frac{1}{4} G \hat{H}_{max} \biggl(\frac{M}{\hat{M}}\biggr)^{\frac{2}{3}} .
\end{equation}
This is the well known result that
the scale of $K$ depends solely on the choice of $M$, for $n$=3 polytropes.

\section{Equilibrium Sequences}

In this section, we summarize the properties of $n$=3
polytropic equilibrium sequences.  The $n'$=3 sequence
constructed by the present authors is discussed in \S 3.1
and the sequences constructed by HTE are discussed in
\S 3.2.

\subsection{$n'$=3}

As mentioned in \S 1,
the $n'$=3 law corresponds to the rotation profile
of a uniformly rotating, spherical $n$=3 polytrope.
In the low-viscosity limit,
a sequence constructed with this law is representative
of the evolution of a SMS that rotates uniformly prior
to its cooling/contraction phase. Thus this sequence
begins with the same starting model as that used in
Baumgarte and Shapiro's study of SMS evolution in
the high-viscosity limit \citep{bash99}.  The $n'$=3 specific
angular momentum distribution is given by \citep{boos73}:
\begin{equation}
j(m)=a_1+a_2\biggl(1-\frac{m(\varpi)}{M}\biggr)^{\alpha_2}
          +a_3\biggl(1-\frac{m(\varpi)}{M}\biggr)^{\alpha_3},
\end{equation}
where $M$ is the total mass of the system, and the numerically
determined constants are $a_1=13.27$, $a_2=163.3$, $a_3=-176.5$,
$\alpha_2=0.2353$, and $\alpha_3=0.2222$.

In Table 1, we list the properties of selected configurations
constructed for the $n'$=3 rotation law.  The data given (in HSCF normalized units, see \S 2.2) include
the axis ratio $R_p/R_e$, polytropic entropy constant $K$ (normalized to 
the entropy constant
of the initial spherical model $K_0$), mass $\hat{M}$, angular momentum
$\hat{J}$, maximum enthalpy $\hat{H}_{max}$, total energy
$\hat{E}=E/G {R_e}^5 {\rho_{max}}^2$,
ratio of kinetic to potential energy $\beta$, ratio of
the centrifugal force $F_c$ to the gravitational force $F_g$
on the equator, and the virial error $VE$.  When the quantity $(F_c/F_g)_{eq}$,
equals
unity, the mass shedding limit $\beta_{shed}$ has been reached.

Figure 1 displays $R_p/R_e$, $\beta$, $(F_c/F_g)_{eq}$, and $E$
as a function of $K/K_0$ for the individual models on this sequence.
Note that $K$, which is a measure of the specific entropy,
decreases monotonically with $R_p/R_e$ along the
sequence. (The ratio $K/K_0$ can be related to time once the cooling
law is specified; this has been treated by Baumgarte and Shapiro 1999
for the case of uniform rotation.)
The energy $E$ also decreases monotonically.
Density contour
plots of selected sequence models are shown in Figure 2.

This sequence
terminates due to mass shedding.
The last spheroidal model we were able to construct had an
axis ratio $R_p/R_e=0.002$ and $(F_c/F_g)_{eq}$=0.99.
We were unable to construct a converged model with
$R_p/R_e$=0.001, as the iteration process oscillated between
models with $(F_c/F_g)$ less than and greater than one.
Thus we conclude that this sequence terminates due to mass
shedding at or very near $R_p/R_e$=0.001, with $\beta_{shed} \sim$ 0.30.
Note that there is no physical effect that forces the mass
shedding limit to occur at an axis ratio of 0.  In
fact, these two points do not coincide on equilibrium
sequences constructed with other equations of state and/or
rotation laws (see, e.g, Hachisu 1986).

Note that no toroidal
branch of this sequence exists.  The HSCF code can construct
tori if the inner radius of the torus $R_i$ is specified instead
of $R_p$. All of the torioidal configurations
we attempted to construct did not converge for this sequence (and
had $(F_c/F_g)_{eq}>1$ during the iteration process). 

Recall that the dynamical bar mode instability is likely to
set in at $\beta_{bar}\lesssim0.27$.  The model with
$\beta \sim \beta_{bar}$=$0.27$ on the $n'$=$3$ sequence
has $R_p/R_e$=$0.004$ and $K/K_0$=$0.381$.

In order to estimate the possible physical scales
of SMS models on this sequence, we must choose the constant values
$M$ and $J$ (see \S 2.2).
For example, say the initial nearly spherical star has
$M$=$10^6 M_{\sun}$, $(R_e)_0$=$10^{17}\,{\rm{cm}}$, and
$\beta_0=(T/|W|)_0$=$10^{-5}$, where a subscript 0 denotes the value is for
the (nearly) spherical star.  
The value of $(R_e)_0$ is chosen such that
the compaction, $(G/c^2) M/R$, for the model that reaches
$\beta \sim \beta_{bar}$=$0.27$ is of the same order of magnitude
as the uniformly rotating critical configuration of Baumgarte
and Shapiro (1999) for gravitational
collapse due to relativistic gravitation. For all
larger values of the product $\beta_0 (R_e)_0$ , an evolved
configuration with $M$=$10^6 M_{\sun}$
will experience a dynamical bar instability prior to reaching the critical
compaction for radial collapse.
The value of $\beta_0$ is quite arbitrary
except for the slow rotation constraint, $\beta_0\ll 1$, necessary
to ensure that the initial cloud is spherical.  Cosmological
considerations and calculations of early structure formation will
ultimately determine the likely distribution of parameters for
primordial SMSs.

Having chosen $M$, $(R_e)_0$ and $\beta_0$, one can
exhibit the scaling of all of the remaining quantities derived
below in terms of these choices, i.e. 
in terms of ($M_6\equiv M/10^6 M_{\sun}$),
($R_{17}\equiv (R_e)_0/10^{17}\,{\rm{cm}}$),
and ($\beta_{-5}\equiv \beta_0/10^{-5}$). 

The choice of $M$, $(R_e)_0$ and $\beta_0$
determines $J$.  To see this, recall that for a uniformly
rotating star $J=(2TI)^{1/2}$, where $I$ is the moment of inertia.
On this sequence, $|\hat W|_0$=$8.96\times 10^{-3}$ and
${\hat I}_0$=$5.82\times 10^{-3}$.  Thus, ${\hat J}_0$=
$3.23\times 10^{-5} \beta_{-5}^{1/2}$.  Equations (8b) and (10)
can then be used to
convert ${\hat J}_0$ to $J$=$1.09\times 10^{61}
[\beta_{-5} R_{17} M_6^3]^{1/2}\,{\rm {g\,cm^2\,s^{-1}}}$.
These choices of $M$ and $J$ for the sequence can be used to dimensionalize
its individual models according to equations (9-12).
For instance, the nearly spherical initial
model has $(\rho_{max})_0$= $2.57\times 10^{-11}
M_6 R_{17}^{-3}\,{\rm {g\,cm^{-3}}}$
and $K_0$=$3.84\times 10^{18}M_{6}^{2/3}
{\rm {g^{-1/3}\,cm^3\,s^{-2}}}$.
The model with $\beta \sim \beta_{bar}=0.27$ and axis ratio 0.004
has $R_e$=$4.14\times 10^{13} \beta_{-5} R_{17}\,{\rm{cm}}$,
$\rho_{max}$=$1.27\times 10^{5}
M_6 [\beta_{-5} R_{17}]^{-3}\,{\rm {g\,cm^{-3}}}$, and
$K$=$1.46\times 10^{18} M_{6}^{2/3}
{\rm {g^{-1/3}\,cm^3\,s^{-2}}}$. 
The model with axis ratio 0.002
has $R_e$=$4.04\times 10^{13} \beta_{-5} R_{17}\,{\rm{cm}}$,
$\rho_{max}$=$4.18\times 10^{5}
M_6 [\beta_{-5} R_{17}]^{-3}\,{\rm {g\,cm^{-3}}}$, and
$K$=$1.20\times 10^{18} M_{6}^{2/3}
{\rm {g^{-1/3}\,cm^3\,s^{-2}}}$. 

\subsection{Hachisu, Tohline, and Eriguchi 1988}

\citet{hach88}, hereafter HTE, used the SCF method to construct 
$n$=3 polytropic sequences with the following parameterized angular
momentum distribution:
\begin{equation}
j(m)=(1+q)(J/M)\bigl[1-(1-m(\varpi)/M)^{1/q}\bigr].
\end{equation}
Here, the index $q$ specifies the rotation law.  Note that the
limiting case of $q$=0 corresponds to the j-constant
rotation law.

The j-constant ($q$=0) sequence HTE constructed is entirely toroidal.
That is, no spheroidal models exist with this rotation law (for any
of the polytropic indices $0 \leq n \leq 3$ they investigated).  Thus
the j-constant sequence is not relevant to the discussion of the
cooling phase of SMS evolution, which begins with a spheroidal configuration.

HTE also constructed sequences with $q$=0.5, 0.7, 1.0, 1.3, and 1.5.  
All of these sequences make continuous transitions between spheroidal
and toroidal configurations at relatively high values of
$\beta_{trans}>$ 0.33, without encountering mass shedding.
Note that if $M$ and $J$ are held constant, the entropy $K$ 
and energy $E$ decrease monotonically
along the spheriodal branch of each of these sequences.  This
confirms that they are suitable representations of the cooling
evolution of SMSs.

From HTE's Figure 7f, one can
estimate that $K_{bar}/K_0\sim 0.4$ for each of these sequences
(where $K_{bar}$ is the value of K for a model with $\beta$=0.27).
Thus the onset of the bar mode instability for HTE's sequences
occurs at approximately the same value of $K/K_0$ as on the
$n'$=$3$ sequence.

\section{Evolutionary Scenarios}

In the low-viscosity limit,
the outcome of the cooling evolution and resultant spin up of
a SMS depends critically on its initial angular momentum
distribution.  Because the nature of the angular momentum distribution
in SMSs is largely unknown, we have presented equilibrium sequences
of SMS models with several different rotation laws in \S 3.  Here
we outline possible scenarios for the outcome of the cooling evolution
of SMSs with low viscosity, based on these sequences.

As mentioned in \S 1, one possible outcome is that a SMS will spin up to
a critical value $\beta_{bar}$, at which it will encounter
the dynamical bar mode instability.  As the bar mode develops,
angular momentum and mass will be transported outwards. The
angular momentum removed will hasten, to some degree, the eventual
collapse of the SMS (and the subsequent possible formation of a SMBH).
A SMS undergoing the bar mode is potentially a strong
source of gravitational radiation because of its highly
nonaxisymmetric structure and hence might be detectable by LISA.

The nonlinear nature of the bar mode instability necessitates
the use of 3D hydrodynamical simulations to precisely determine
$\beta_{bar}$ for each equilibrium sequence.  However, previous
linear and nonlinear stability analyses indicate
that $0.14\lesssim\beta_{bar}\lesssim 0.27$
for a wide range of polytropic indices and rotation laws
\citep{shte83,duto85,imam85,mana85,hach88,toha90,pick96,toma98,idp00}.
The lower limit, $0.14$, may be most appropriate
for spheroids with off-center density maxima and tori \citep{hach88,
toha90}.

The nonlinear hydrodynamics study of \citet{pick96} indicates that
$\beta_{bar}$ may be less than $0.27$ for rotation laws that
place a significant amount of angular momentum in equatorial mass
elements.  Their results predict that the $m$=2 stability limit may
be less than $\sim 0.20$ for models with the $n'$=3 rotation law.
Their simulations, of $n$=1.5 polytropes, also suggest that a
one-armed spiral, $m$=1 mode may become increasingly dominant over
the $m$=2 mode as the equatorial concentration of angular momentum increases.
Note that the grid resolutions used in \citet{pick96}
were likely not sufficient to accurately model the development of
instabilities in models with these extreme differential rotation
laws \citep{toma98}.  However, the results of \citet{pick96} and
the linear and nonlinear stability analyses of \citet{toma98}
confirm that $0.27$ is likely to be an approximate upper limit
to $\beta_{bar}$, for a variety of polytropic indices and rotation
laws.  The analysis presented in the remainder of this
manuscript assumes that the $m$=2 bar mode is the dominant mode and that
$\beta_{bar}\lesssim 0.27$.

The new $n'$=3 sequence constructed by the present authors and the
sequences constructed by HTE that are considered in this work
indicate that a bar mode phase is probable for SMSs with a wide range of 
differential rotation laws.  Indeed, HTE's $q$=0.1-1.5 (see equation 14)
sequences do not terminate due to mass shedding.  Each of these sequences makes
a continuous transition from spheroidal to toroidal configurations
at values of $\beta_{trans}>0.33>0.27\gtrsim\beta_{bar}$.
Thus, $n$=3 models with these
$q$-indexed laws would likely encounter the bar mode as spheroids.

The spheroidal $n'$=3 sequence terminates due to mass shedding
at $\beta_{shed}\gtrsim 0.30$.  Because $\beta_{shed}>0.27$, we
expect that models with this rotation law become dynamically
unstable to the bar mode near $R_p/R_e \sim 0.004$ (this model is
marked with a solid dot in Figure 1), before reaching
the mass shedding limit. 

Note that even if the actual value of $\beta_{bar}$ is less than
$0.27$, the qualitative nature of our results would not change.
That is, the sequences examined here would still have models that
are unstable to the bar mode.  In addition, the quantitative estimates
of the characteristics of the gravitational radiation emission
presented in \S 5 would only change by a numerical factor of order
1-10, even if $\beta_{bar}$ were as low as $0.22$.

3D hydrodynamical simulations are needed
to properly study the evolution of rapidly rotating stars
that encounter the bar mode instability.  To date, we are unaware
of any published hydrodynamical studies of the bar mode
in $n$=3.0 polytropes.  However, many authors have simulated this
instability in polytropes with lower $n$ and various rotation laws
(e.g., Tohline, Durisen, and McCollough 1985; Pickett, Durisen, and
Davis 1996; Imamura, Durisen, and Pickett 2000;
New, Centrella, and Tohline 2000; Brown 2000).
All of these simulations agree
on the initial phase of the instability, during which the initially
axisymmetric object deforms into a bar shape through which spiral
arms are shed. Recent Newtonian simulations indicate that the
long-term outcome of the 
bar instability in an $n$=1.5 polytrope (with an initial Maclaurin rotation
law) is a persistent bar-like structure that emits
gravitational radiation over many rotation periods \citep{nct00,brow00}
(the persistent nature of the bar structure was recently confirmed
with the post-Newtonian simulations of Saijo et al. 2000).
Simulations relevant to the dynamical bar mode instability
have recently been performed for $n$=1 polytropes in full 3D general
relativity by \citet{shib00}.

\section{Conclusions and Future Work}

The thermal emission of a rotating supermassive star causes it
to contract and spin up.  If the viscosity and internal magnetic
fields are weak, the
SMS will rotate differentially during its cooling and contraction phase.
We have examined models of the quasistatic
cooling evolution of SMSs, represented by equilibrium sequences
with six differential rotation laws.

The evolution of a cooling SMS that is initially nearly spherical,
with very low, uniform rotation can be represented by the $n'$=3
sequence we constructed.
This sequence terminates due to mass--shedding at $\beta_{shed}\gtrsim 0.30$.
This $\beta_{shed}$ exceeds the value $0.27$, which
is a likely upper limit for the onset of
the rotationally
induced, dynamical bar instability.  Thus it appears likely that a star
with this rotation law would become unstable to the bar mode prior
to reaching the mass shedding  limit.

We have also considered sequences constructed by HTE, in order to
follow the evolution of SMSs with five different initial
differential rotation laws.  No mass--shedding limits exist
on these sequences.  Thus, a cooling SMS characterized by one of these
rotation profiles will also encounter the dynamical bar mode.

This mode will transport mass and angular momentum outward and thus may
hasten the onset of collapse and possible SMBH formation
\citep{tdm85,pick96,idp00,nct00,brow00}. 
Because of their nonaxisymmetric structure, SMSs undergoing the
bar mode could be strong sources of quasi-periodic gravitational radiation
for detectors like LISA, even before collapse.

The frequency of these quasiperiodic gravitational waves can be
estimated from the expected bar rotation rate $\Omega_{bar}$.
The model on our $n'$=$3$ sequence with $\beta\sim\beta_{bar}$=$0.27$
has a central rotation rate $\Omega_{c}$=$1.02\times 10^{-1}
{\rm Hz}\, [M_6 \beta_{-5}^{-3} R_{17}^{-3}]^{1/2}$.  In previous hydrodynamics simulations of the bar mode
instability \citep{nct00}, $\Omega_{bar}$ was $\sim 0.4 \Omega_{c}$.  With
this relation between $\Omega_{bar}$ and $\Omega_{c}$, the gravitational
wave frequency $f_{GW}$ can be estimated to be
\begin{eqnarray}
f_{GW} &=& 2 f_{bar} = 2 \frac{\Omega_{bar}}{2\pi} \sim \frac{0.4}{\pi}\Omega_{c} \nonumber \\
&\sim& 1\times 10^{-2} {\rm Hz}\, [M_6 \beta_{-5}^{-3} R_{17}^{-3}]^{\frac{1}{2}}.
\end{eqnarray}
For a SMS of $10^6 M_{\sun}$, which begins as a slowly rotating star of radius
$10^{17} {\rm cm}$ with $\beta_0$=$10^{-5}$, this yields a frequency of
$1\times 10^{-2} {\rm Hz}$. This frequency
is in the range in which LISA is expected to be most sensitive,
$10^{-4}$-$1 {\rm Hz}$
(see, e.g., Thorne 1995, Folkner 1998).
The choice $\beta_0$=$2\times 10^{-4}/R_{17}$ is the maximum value of $\beta_0$ 
for which $f_{GW}$ (=$10^{-4} {\rm Hz}$) is still in LISA's
range of sensitivity.\footnote{We note that \citet{peeb69}
has discussed the spin-up
of primordial masses by tidal torquing and finds that $J \sim M R^2$.
Naively scaling his results (derived for $M \sim 10^{11}
M_{\sun}$ configurations)
to our regime, we find that $\beta_0 \sim 10^{-4}$, so that the resulting
frequency of gravitational radiation would lie near the lower
limit of LISA's range.}

The strength of the gravitational wave signal can be estimated roughly to
be
\begin{eqnarray}
h &\sim& \frac{G}{c^4} \frac{\ddot{Q}}{d} \sim \frac{G}{c^4}
\frac{M R_{bar}^2 f_{bar}^2}{d} \nonumber \\
  &\sim& 4 \times 10^{-15} \biggl(\frac{d}{1\, {\rm Gpc}}\biggr)^{-1}
M_6^2 \beta_{-5}^{-1} R_{17}^{-1},
\end{eqnarray}
where $\ddot{Q}$ is the second time derivative of the star's
quadrupole moment and $d$ is the distance, which we scale to 1 Gpc
(the Hubble distance is $\sim$ 3 Gpc).  Here we have used $R_{bar}
\sim R_e$=$4.14\times 10^{13} \beta_{-5} R_{17}\,{\rm{cm}}$ for the
$n'$=$3$ model that reaches the point $\beta\sim\beta_{bar}$.

As mentioned previously, recent simulations of the bar instability
indicate that the outcome is a long-lived bar-like structure
\citep {nct00, brow00, saij00}.  The bar will decay on a secular
timescale due to dissipative effects.  For differentially rotating
SMSs, the largest source of dissipation will be gravitational radiation.
The gravitational radiation damping timescale $\tau_{GW}$
is approximately
\begin{equation}
\tau_{GW} \sim \frac{T}{\bigl(\frac{dE}{dt}\bigr)_{GW}},
\end{equation}
where $T$ is the rotational kinetic energy and
$(dE/dt)_{GW}$ is the rate at which gravitational radiation carries
energy away from the system.  Recall that $T=\beta |W|$.
The gravitational potential energy $|W| \sim GM^2/R_{bar}$.  Thus,
\begin{equation}
T=\beta \frac{G M^2}{R}.
\end{equation}
The radiation rate can be estimated as  \citep{shte83}
\begin{equation}
\bigl(\frac{dE}{dt}\bigr)_{GW} \sim \frac{G}{c^5} \frac{M}{R^2} v^6.
\end{equation}
In this case the characteristic velocity of the system is
$v=(\beta G M/R)^{1/2}.$  Substitution of equations (18) and (19)
into equation (17) yields
\begin{eqnarray}
\tau_{GW} &\sim& \frac{c^5}{G^3} \frac{R_{bar}^{4}}{\beta^2 M^3} \nonumber \\
          &\sim& 1 \times 10^4 yrs  M_6^{-3} [\beta_{-5} R_{17}]^4.
\end{eqnarray}

The number of cycles $\cal{N}$ for which the signal will persist is
\begin{eqnarray}
\cal{N}&=&\tau_{GW} f_{GW} \nonumber \\
 &=& 4 \times 10^9 [M_6 \beta_{-5} R_{17}]^{-5/2}.
\end{eqnarray}
The quasiperiodicity of such a signal will assist in its detection
\citep{schutz97}.

The fraction of the mass $(dM/M)_{GW}$ radiated via gravitational radiation over
the interval $\tau_{GW}$ can be estimated as
\begin{eqnarray}
\biggl(\frac{dM}{M}\biggr)_{GW} &\sim& \tau_{GW}
 \bigl(\frac{dE}{dt}\bigr)_{GW} \frac{1}{Mc^2} \nonumber \\
        &\sim& \frac{G}{c^2} \frac{\beta M}{R} \nonumber \\
        &\sim& 1 \times 10^{-3} M_6 [\beta_{-5} R_{17}]^{-1}.
\end{eqnarray} 

It appears that SMSs with a wide
range of differential rotation laws
may become unstable to the development of the dynamical bar mode.
However, linear and nonlinear stability analyses are needed to precisely
determine the onset of the $m$=2 dynamical instability
for the equilibrium sequences considered here.  Such analyses are
also needed to determine the relative importance of various unstable
nonaxisymmetric modes in SMS models (as a dominant $m$=1 mode would 
change the characteristics of the gravitational radiation emission).
In addition, three-dimensional hydrodynamical simulations
are necessary to study the mass and angular momentum redistribution
induced by these nonaxisymmetric instabilities, to
compute the gravitational waveforms
emitted, and to determine the final fate of the star.

Future investigations involving hydrodynamical simulations of
SMS models would benefit from improved knowledge of initial conditions,
such as an appropriate value for $\beta_0$.  These appropriate
initial conditions could be determined from studies of large-scale
structure and cosmology.

Most importantly, studies are required to assess the roles of viscosity
and magnetic fields in rotating SMSs to judge whether they are sufficient
to drive these configurations to uniform rotation prior to bar instability.



\acknowledgments
We thank Joel Tohline for helpful conversations.
This work has been supported in part by NSF Grants AST 96-18524
and PHY 99-02833 and NASA Grants NAG5-7152 and NAG5-8418 to the
University of Illinois at Urbana-Champaign. A portion of this work
was performed under auspices of the U.S. Department of Energy
by Los Alamos National Laboratory under contract W-7405-ENG-36.

\clearpage



\begin{figure}
\plotone{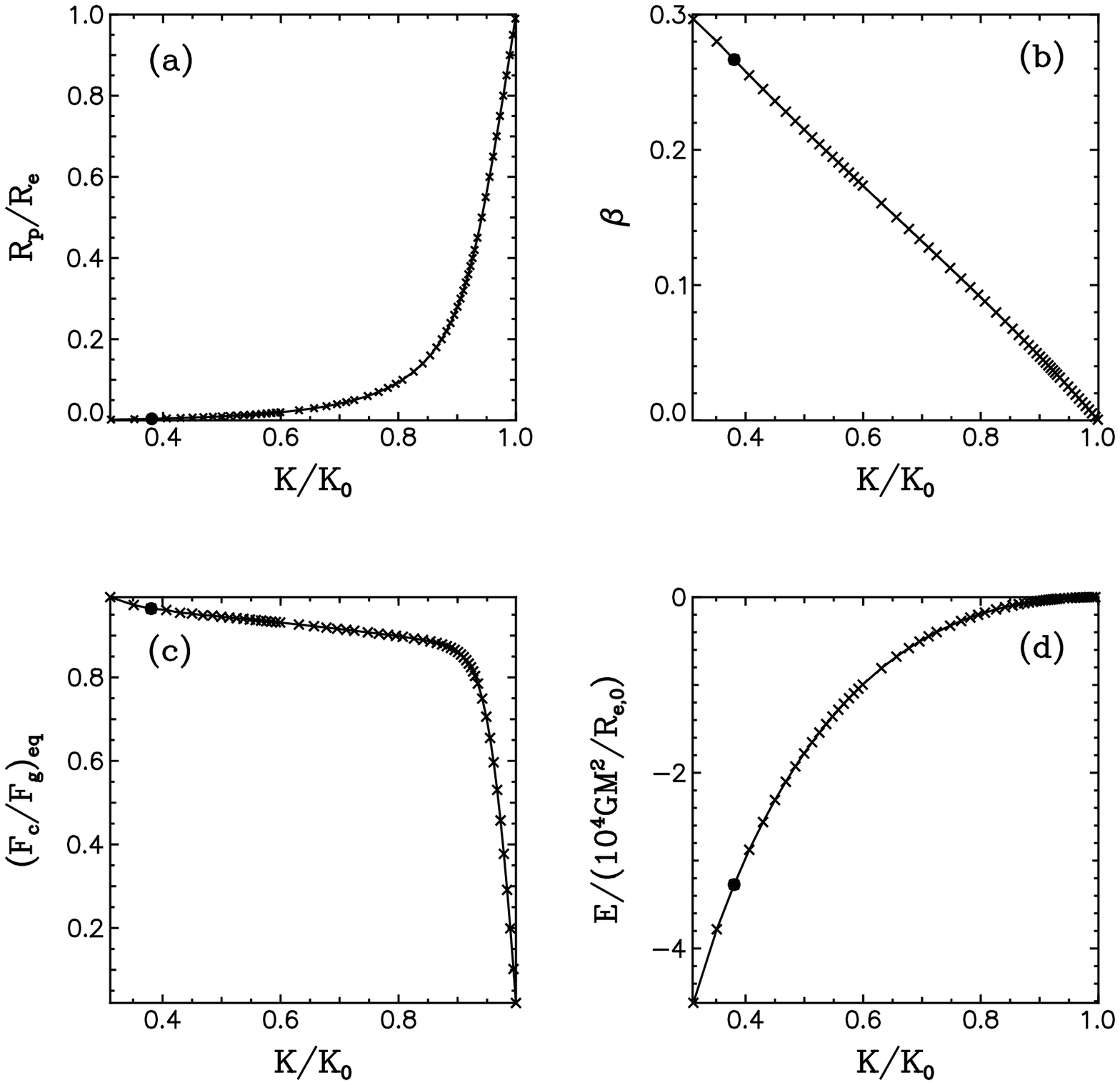}
\caption{Properties of individual models on the
$n'$=3 equilibrium sequence are displayed as a
function of the ratio $K/K_0$, where $K_0$ is the entropy constant
for the initial, spherical model:  (a) $R_p/R_e$, axis ratio,
(b) $\beta$, ratio of kinetic to gravitational
potential energy; (c) $(F_c/F_g)_{eq}$, equatorial ratio of centrifugal
to gravitational force; (d) $E$, total
energy, in units of $G M^2/(R_e)_0$.
The cooling evolution of a star that is represented by this sequence
begins with the spherical model, for which $K/K_0$=1, and
contracts along the sequence characterized by
monotonically decreasing $K/K_0$. Quasistationary evolution proceeds until
the bar mode instability is encountered at $\beta_{bar}\sim 0.27$.  The
model for which $\beta$=$\beta_{bar}\sim 0.27$
is marked with a solid dot in each
frame.}
\end{figure}

\begin{figure}
\plotone{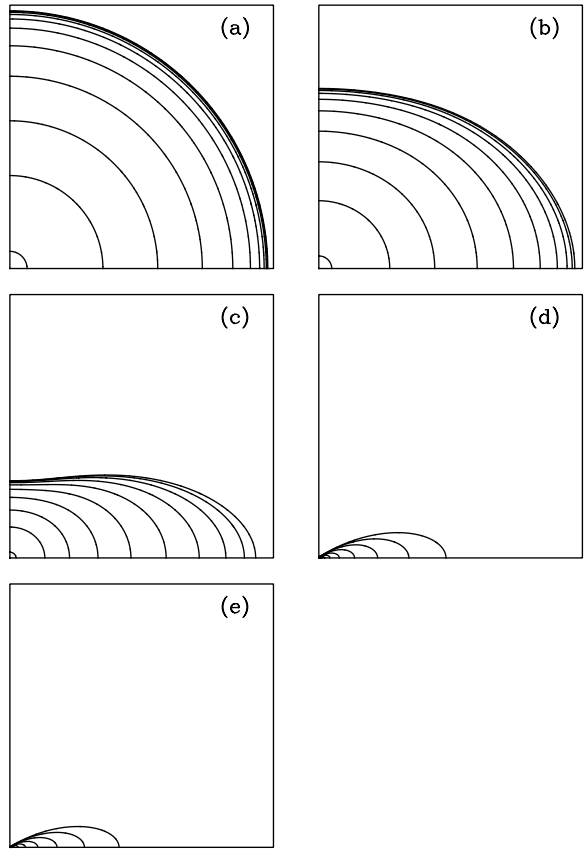}
\caption{Density contours of selected models on the
$n'$=3 equilibrium sequence are shown in the
$(x>0,z>0)$ plane.  The maximum density $\rho_{max}$ is normalized
to unity.  The highest density contour level is 0.9; subsequent
contour levels range from $10^{-1}$ to $10^{-10}$ and are
separated by a decade.  The axis ratios $R_p/R_e$ of the models displayed
are (a) 1.00; (b) 0.700; (c) 0.300; (d) 0.004; (e) 0.002.
The model with $R_p/R_e$=0.004 shown in (d) has $\beta$=$\beta_{bar}
\sim 0.27$.}
\end{figure}



\clearpage

\begin{deluxetable}{ccrrrrrcr}
\footnotesize
\tablecaption{$n'$=3 sequence. \label{tbl-3}}
\tablewidth{0pt}
\tablehead{
\colhead{$R_p/R_e$} & \colhead{$K/K_0$} & \colhead{$\hat{M}$} &
\colhead{$\hat{J}$} &
\colhead{$\hat{H}_{max}$} &
\colhead{$\hat{E}$} & \colhead{$\beta$} & \colhead{$(F_c/F_g)_{eq}$} &
\colhead{$VE$} 
}
\startdata
1.000 & 1.000 & 7.73E-2 & 0.00    & 0.264 &  1.68E-8 & 0.00    &0.000&1.87E-6 \\
0.900 & 0.989 & 6.09E-2 & 5.06E-4 & 0.223 & -3.16E-5 & 5.28E-3 &0.199&2.10E-6 \\
0.800 & 0.978 & 4.62E-2 & 4.61E-4 & 0.183 & -4.01E-5 & 1.07E-2 &0.378&2.58E-6 \\
0.700 & 0.967 & 3.35E-2 & 3.37E-4 & 0.146 & -3.53E-5 & 1.62E-2 &0.531&3.46E-6 \\
0.600 & 0.955 & 2.28E-2 & 2.11E-4 & 0.112 & -2.52E-5 & 2.19E-2 &0.655&3.94E-6 \\
0.500 & 0.942 & 1.44E-2 & 1.13E-4 &8.11E-2& -1.49E-5 & 2.82E-2 &0.749&4.67E-6 \\
0.400 & 0.926 & 8.13E-3 & 4.99E-5 &5.45E-2& -7.13E-6 & 3.54E-2 &0.814&6.67E-6 \\
0.300 & 0.906 & 3.88E-3 & 1.69E-5 &3.26E-2& -2.60E-6 & 4.48E-2 &0.854&1.25E-5 \\
0.200 & 0.874 & 1.39E-3 & 3.66E-6 &1.58E-2& -6.04E-7 & 5.91E-2 &0.879&2.56E-5 \\
0.100 & 0.807 & 2.52E-4 & 2.86E-7 &4.69E-3& -4.95E-8 & 8.79E-2 &0.897&7.84E-5 \\
0.050 & 0.725 & 4.90E-5 & 2.47E-8 &1.41E-3& -4.17E-9 & 0.122   &0.911&2.28E-4 \\
0.040 & 0.696 & 2.94E-5 & 1.15E-8 &9.64E-4& -1.89E-9 & 0.134   &0.916&3.17E-4 \\
0.030 & 0.657 & 1.53E-5 & 4.34E-9 &5.90E-4& -6.86E-10& 0.150   &0.922&4.87E-4 \\
0.020 & 0.599 & 6.24E-6 & 1.13E-9 &2.96E-4& -1.66E-10& 0.174   &0.931&8.97E-4 \\
0.010 & 0.500 & 1.41E-6 & 1.22E-10&9.14E-5& -1.49E-11& 0.215   &0.945&2.40E-3 \\
0.008 & 0.468 & 8.84E-7 & 6.09E-11&6.28E-5& -6.90E-12& 0.228   &0.948&3.18E-3 \\
0.006 & 0.430 & 4.90E-7 & 2.52E-11&3.89E-5& -2.57E-12& 0.245   &0.954&5.29E-3 \\
0.005 & 0.406 & 3.41E-7 & 1.47E-11&2.88E-5& -1.39E-12& 0.255   &0.961&7.23E-3 \\
0.004 & 0.381 & 2.21E-7 & 7.67E-12&2.02E-5& -6.62E-13& 0.267   &0.964&9.85E-3 \\
0.003 & 0.351 & 1.31E-7 & 3.53E-12&1.32E-5& -2.68E-13& 0.280   &0.973&1.50E-2 \\
0.002 & 0.311 & 7.20E-8 & 1.44E-12&7.84E-6& -9.66E-14& 0.297   &0.992&1.54E-2 
\enddata
\end{deluxetable}






\end{document}